\documentclass[sn-aps]{sn-jnl} 

\usepackage{graphicx}%
\usepackage{multirow}%
\usepackage{amsmath,amssymb,amsfonts}%
\usepackage{amsthm}%
\usepackage{mathrsfs}%
\usepackage[title]{appendix}%
\usepackage{xcolor}%
\usepackage{textcomp}%
\usepackage{manyfoot}%
\usepackage{booktabs}%
\usepackage{algorithm}%
\usepackage{algorithmicx}%
\usepackage{comment}
\usepackage{algpseudocode}%
\usepackage{listings}%
\usepackage[section]{placeins}
\usepackage{booktabs}%
\usepackage{ulem}
\usepackage{caption,subcaption,tikz}

\usetikzlibrary{arrows.meta, shapes.geometric, arrows, positioning, shadows}
\DeclareCaptionFormat{suboverlay}{\gdef\subcapoverlay{(\thesubfigure) #3\par}}
\DeclareCaptionStyle{suboverlay}{format=suboverlay}

\begin{document}

\title{Data-Driven Analysis of Droplet Morphology in Inkjet Systems: Toward Generating Stable Single-Drop Regimes}


\author*[1]{\fnm{Ali R.} \sur{Hashemi}}\email{ahashemi@cimne.upc.edu}

\author[1]{\fnm{Angela M.} \sur{Ares de Parga-Regalado} } \email{aares@cimne.upc.edu}

\author*[1,2]{\fnm{Pavel B.} \sur{Ryzhakov} } \email{pavel.ryzhakov@upc.edu}

\affil[1]{\textit{\orgname{Centre Internacional de Mètodes Numérics en Enginyeria (CIMNE)}, \orgaddress{\street{ C/Gran Capitán s/n}, \city{Campus Nord UPC}, \postcode{08034}, \state{Barcelona}, \country{Spain}}}}

\affil[2]{\textit{\orgdiv{Escola Tècnica Superior d’Enginyers de Camins, Canals i Port}, \orgname{Universitat Politècnica de Catalunya - BarcelonaTech (UPC)}, \orgaddress{\street{ C/Jordi Girona 1}, \city{Campus Nord UPC}, \postcode{8034}, \state{Barcelona}, \country{Spain}}}}


\abstract{\color{black} The growing demand for new microelectronic devices and pharmaceutical advancements has heightened interest in inkjet printing as a means of high-precision manufacturing technique. This study leverages data-driven analyses to optimize droplet generation processes in a drop-on-demand dispensing system. A three-voltage pulse scheme was employed to produce droplets, with high-resolution images captured and processed to extract geometric features of the principal droplet. This resulted in a comprehensive, openly published dataset, along with a detailed, reproducible image processing pipeline. By analyzing this data, we identified key operational parameters and established correlations between inputs and outputs, providing insights into consistent single-droplet generation. These findings offer practical guidelines for controlling droplet morphology and advancing applications in inkjet printing\color{black}}




\maketitle

\section{Introduction}

In the past decade, inkjet printing (IJP) technology has expanded well beyond its traditional applications. It is now utilized in the printing of high-precision optical and electronic components, as well as in biomedical and pharmaceutical fields, particularly for drug administration \cite{vespini_forward_2016,zhan_inkjet-printed_2017,ito_fabrication_2021,carou-senra_inkjet_2024}. Many of these applications demand precise control over ink deposition, making drop-on-demand (DoD) devices the preferred choice due to their high accuracy, efficiency, and suitability for automation \cite{castrejon-pita_future_2013}. Two primary types of DoD print heads dominate: thermal and piezoelectric. While thermal print heads are inexpensive and commonly used in household printers, they present limitations, particularly in pharmaceutical applications where thermolabile materials are involved \cite{carou-senra_inkjet_2024}. On the other hand, piezoelectric devices, which deform through voltage pulses, offer greater versatility across various applications \cite{shah_classifications_2021}. In addition to these two types, devices built on other  principles, such as electrohydrodynamics, are emerging, striving to improve efficiency, reduce costs, and enhance precision \cite{shadloo_sph_2013,narvaez-munoz_computational_2024}.

Although significant improvements have stemmed from experimental studies, these experiments are often costly due to the many variables that need to be controlled \cite{dong_visualization_2006,yang_drop--demand_2018}. As a result, regardless of the type of print head, there is a growing need for virtual analyses in microfluidic processes \cite{hashemi_enriched_2020,hashemi_toward_2021}. These virtual analyses allow for the exploration of parameter variations to achieve desired outcomes more efficiently. For instance, while DoD micro-dispensers have produced high-quality results in printing electronic components, challenges related to printability and pattern resolution persist \cite{Lemarchand_challenges_2022}. Moreover, the production of microcapsules through droplet microfluidics has gained prominence, with droplet shape playing a crucial role in biomedicine applications \cite{duran_microcapsule_2022}. Consequently, numerical simulations have become integral to optimizing inkjet systems, focusing on droplet morphology and print resolution \cite{hu_morphology_2022,xia_droplet_2024,lohse_fundamental_2022}. Given this context, controlling droplet geometry through operational parameters has emerged as a critical area of research.

Each kind of inkjet dispensers operates according to a unique set of parameters. Piezoelectric devices, in particular, are defined by voltage pulse characteristics such as amplitude, duration, and pulse timing. The shape of the voltage pulse is also essential in determining the final droplet geometry \cite{shah_actuating_2020,shah_classifications_2021}. Research has largely focused on how droplet formation responds to changes in voltage signals, with attention to physical properties like density, viscosity, surface tension, and droplet size \cite{bogy_experimental_1984,malloggi_electrowetting_2008,jang_influence_2009,shin_control_2011,shin_shape_2014,gong_characterization_2023}. However, despite these efforts, there remain significant opportunities to develop more efficient tools for real-time decision-making and optimization. In this regard, data-driven methods and artificial intelligence offer promising solutions for improving inkjet system performance \cite{kim_design_2022,dong_development_2023,phung_machine_2023,ryzhakov_international_2024,takenaka_classification_2024}.

With the current ability to handle large datasets, it is now possible to directly analyze the relationships between operational parameters of droplet dispensers and the morphological attributes of the droplets produced. This study aims to do just that by examining the voltage pulse schemes of a piezoelectric dispenser and predicting droplet morphology based on these schemes.

To this end, a substantial collection of images was captured from the output droplets on the inkjet dispenser, allowing for the extraction of various geometric attributes. These images were post-processed and labeled according to both the operational parameters of a triple voltage pulse scheme and the geometric features of the principal droplet. A detailed analysis of the relationships between input parameters and output droplet characteristics followed, leading to a classification of droplet morphology based on the operational regime.

The remainder of this paper is structured as follows: Section \ref{Ex} describes the experimental setup used for droplet generation and the methodology employed for image processing and geometric feature extraction. Section \ref{Data} presents a comprehensive examination of the data, including distribution analysis and the handling of outliers. Additionally, this section delves into the geometric outputs that define droplet morphology. Section \ref{Phys} offers a physical interpretation of the data, highlighting the correlation between voltage pulse schemes and droplet shape attributes. Finally, in Section \ref{Con}, we summarize the key findings and provide an outlook on future research directions        

\section{Methods}\label{Ex}

\subsection{Experimental setup}
\begin{figure}[h]
    \centering
    \includegraphics[width=0.6\linewidth]{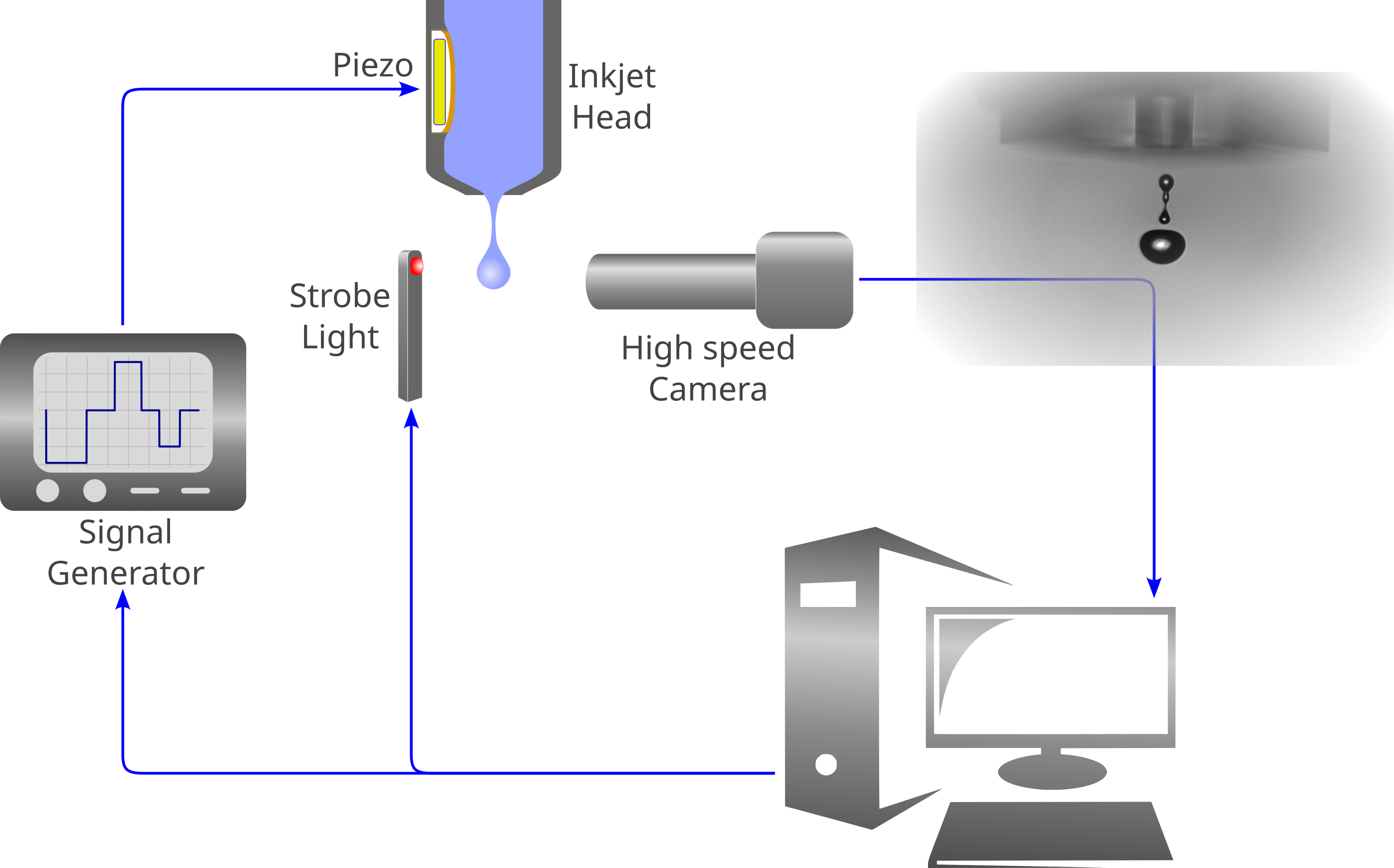}
    \caption{Schematic representation of the experimental setup for droplet generation and image capture, utilizing a piezoelectric dispenser, signal generator, stroboscopic lighting, and high-speed camera.}
    \label{fig:exp-setup}
\end{figure}

In this study, the Microdrop dispensing system MD-E-5000 was used, which consists of a piezoelectric dispenser and a control unit responsible for generating input signals to the dispenser, as well as managing the stroboscopic light source. A high-speed camera was employed to capture images of droplets generated by varying input parameters. Due to the widespread use of alcohol-based liquids in inkjet devices, ranging from conventional printing to bio-printing and additive manufacturing, ethanol with a purity of $99\%$ was chosen as the working fluid. The experimental setup is schematically illustrated in Fig.~\ref{fig:exp-setup}.

\begin{figure}[h]
    \centering
    \includegraphics[width=0.45\textwidth]{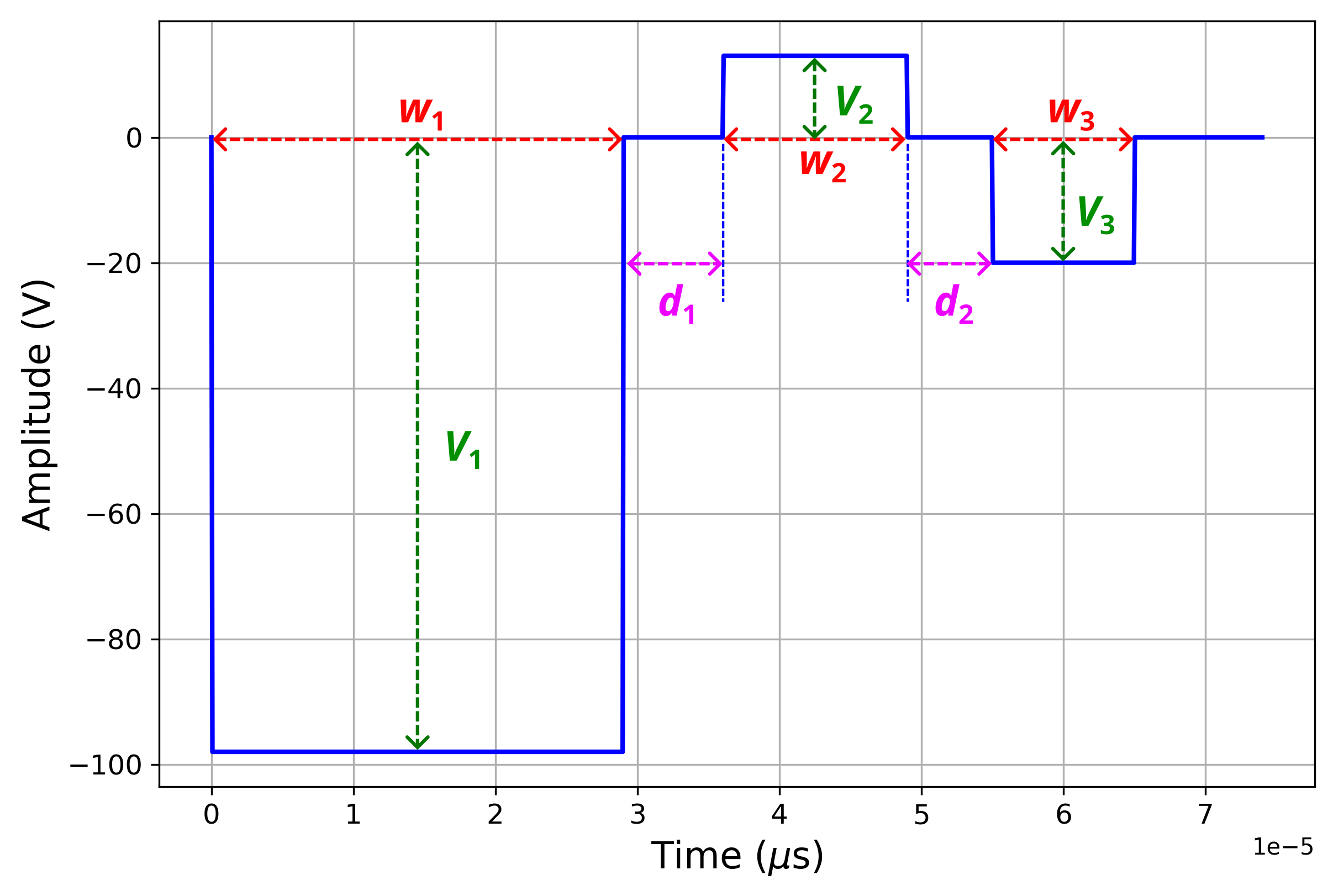}
    \caption{Input signal characteristics for a sample configuration ($V_1=-98$V, $V_2=13$V, $V_3=-20$V, $w_1=29\mu$s, $w_2=13\mu$s, $w_3=10\mu$s, $d_1=7\mu$s and $d_2=6\mu$s).}
    \label{fig:input_signal}
\end{figure}

The piezoelectric dispenser converts voltage pulses into mechanical pulses to form droplets \cite{sakai_dynamics_2000}. Rectangular pulses were used for this study, as they are the preferred choice recommended by the manufacturer to ensure the reliable functionality of the dispenser. The amplitude of each voltage pulse is translated into acoustic waves, generated by the vibrations of the piezoelectric material. The input signal consists of three consecutive pulses, each independently adjustable by eight key parameters: the amplitudes of the three pulses ($V_1$, $V_2$, $V_3$), the widths of the pulses ($w_1$, $w_2$, $w_3$), and the time intervals between the pulses ($d_1$, $d_2$). Figure \ref{fig:input_signal} shows the signal shape for a specific configuration, and Table \ref{tab:inputs} presents the range of input parameters explored in this study. We note that the values of $V_1$ and $V_3$ used in this work are always negative. 

\begin{table}[h]
    \centering \footnotesize
    \caption{Range of input parameters used in the experiments.}
    \label{tab:inputs}
    \begin{tabular}{@{}lcccccccc@{}}
        \toprule
        &$V_1$(V) & $V_2$(V) & $V_3$(V) & $w_1(\mu$s) & $w_2(\mu$s) & $w_3(\mu$s) & $d_1(\mu$s) & $d_2(\mu$s) \\
        \midrule
        max & -82 & 14 & 0 & 35 & 15 & 31 & 7 & 7 \\
        min & -98 & 12 & -30 & 20 & 10 & 1 & 5 & 5 \\
        step & 4 & 1 & 5 & 3 & 1 & 3 & 1 & 1 \\
        \botrule
    \end{tabular}
\end{table}

The first pulse generates an acoustic wave that pushes liquid from the nozzle, forming the primary droplet. The second pulse creates a negative pressure wave, pulling the liquid back, thus reducing the droplet’s velocity and forming its tail. This pulse correlates with the formation of satellite droplets and liquid threads, depending on its properties \cite{shah_classifications_2021}. The third pulse assists in detaching the droplet from the nozzle, playing a significant role when conditions for droplet formation are not fully achieved by the previous pulses. The time intervals between pulses ($d_1$, $d_2$) influence the interaction between acoustic waves and can cause disturbances in the droplet formation process \cite{shin_control_2011}.

\subsection{Data Collection and Open Data}
All images captured by the high-speed camera are stored in JPEG format with filenames that include the input signal parameters, ensuring traceability and organization. This extensive dataset is published on Zenodo, a widely used open-access repository developed by CERN and OpenAIRE, under an open data license. The dataset~\cite{hashemi_inkjet_2024} adheres to FAIR principles by providing a unique DOI for findability, being accessible in standard formats (JPEG for images and CSV for data), and allowing interoperability and reuse by researchers worldwide.

Additionally, the input parameters and extracted geometrical characteristics of the droplets are made available in CSV format, facilitating easy analysis. The Python code for controlling the experimental setup, handling data, and processing images is openly accessible on our GitHub repository~\cite{hashemi_inkjetdroplet_2024}, ensuring transparency and reproducibility.

\subsection{Image processing}
The OpenCV library was used to process the high-resolution images of inkjet droplets and accurately extract droplets characteristics. The image processing pipeline includes the following key steps:

\textbf{Image cropping:} The high-resolution images were cropped to focus on the region containing the droplet, removing irrelevant areas and reducing computational costs.

\textbf{Image enhancement:} After converting to gray-scale, Contrast Limited Adaptive Histogram Equalization (CLAHE) was applied to improve contrast while avoiding noise amplification.

\textbf{Noise reduction:} A Gaussian blur filter was applied to reduce noise and smooth the image, eliminating small fluctuations caused by sensor noise or other artifacts.

\textbf{Binarization:} Grayscale images were converted to binary using thresholding, separating droplets (foreground) from the background.

\textbf{Contour extraction:} Iso-intensity contours were extracted from the binarized images, identifying droplet boundaries and bright spots caused by light reflection. The hierarchy of contours was analyzed to distinguish between parent contours (droplet boundaries) and child contours (bright spots).

\textbf{Droplet features:} Ellipses were fitted to the contours, providing information about the droplet's shape and size through the semi-axes ($a$ and $b$).

\textbf{Non-elliptic droplets:} Some droplets exhibited tails or irregular shapes. In such cases, the number of child contours associated with a parent contour was checked to identify non-elliptical shapes. For droplets with tails, morphological transformations (distance transform and dilation) were applied to isolate the main droplet.

\color{black}
The image processing pipeline is fully automated, processing the entire database of raw images and generating a dataset of extracted features without manual intervention. Manual input is limited to parameter configuration, optimized through trial-and-error on a small subset of images, as consistent imaging conditions ensured reliability without requiring adaptive optimization.

Verification is performed via manual inspection of random image samples and a clustering-based data cleaning method to identify and correct outliers or pipeline errors. Details of this robust cleaning approach are provided in Section~\ref{data_cleaning}.
\color{black}

\begin{figure}
    \centering
    \includegraphics[width=0.5\textwidth]{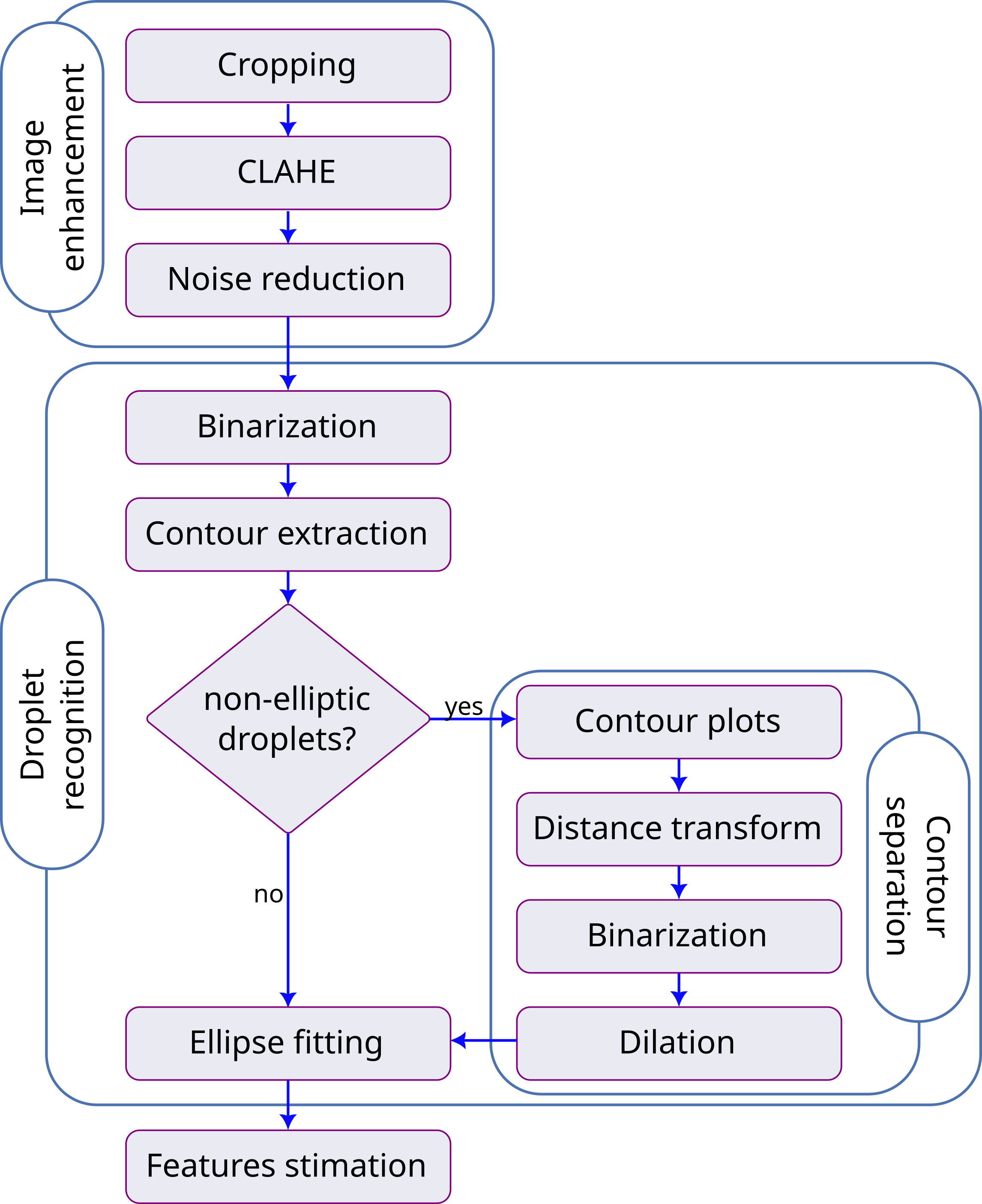}
    \caption{Flowchart of the droplet recognition process.}
    \label{fig:image_process}
\end{figure}

The flowchart in Fig~\ref{fig:image_process} summarizes the image processing pipeline. 

\section{Data Analysis}\label{Data}
The collection of images obtained in this work~\cite{hashemi_inkjet_2024} is valuable as it encompasses a wide range of droplet attributes captured at a time-lapse of $150$ $\mu$s. This time corresponds to the distance between the print-head and substrate in IJP devices. Therefore, the database can be used to infer physical characteristics by replicating IJP processes.

\begin{figure}[!h]
    \centering
    \includegraphics[width=0.6\textwidth]{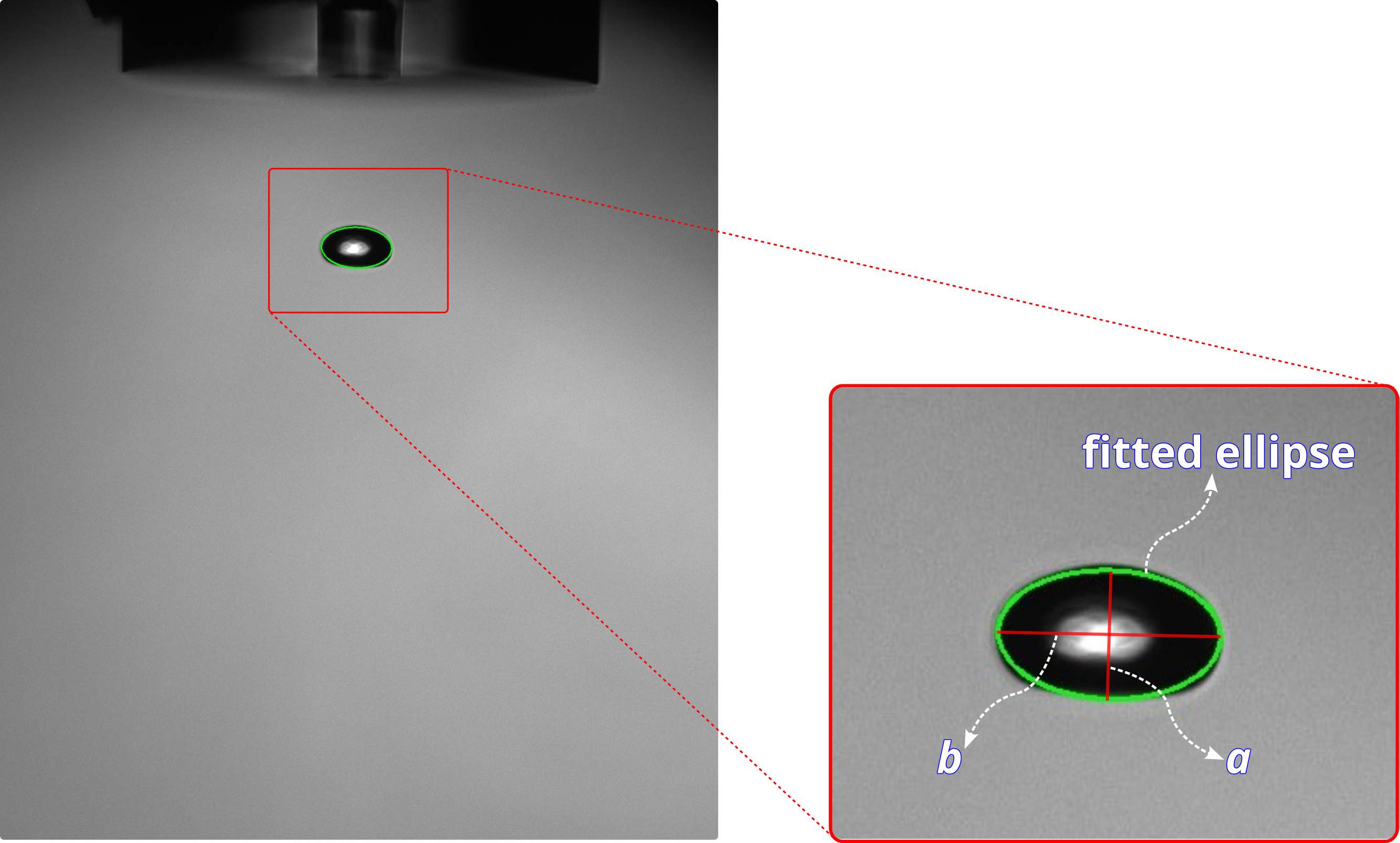}
    \caption{Visualization of the semi-axes $a$ and $b$ that define the ellipse enclosing the principal droplet.}
    \label{fig:ellipse_a_b}
\end{figure}

The established database contains more than 300,000 samples, providing the operating parameters for each image and the geometric components characterizing the ellipses fitted to the principal droplet. These components include the ellipse semi-axes lengths, $a$ and $b$ (see Fig.~\ref{fig:ellipse_a_b}), the ellipse centroids, and the $\theta$ angle used to rotate the images to fit the ellipses. Additionally, other ellipses were fitted to capture secondary droplets structures. Thus, the database also includes the number of fitted ellipses $n$ for each image.

This work focuses on three output parameters: $a$, $b$, and $n$, as this triplet represents the shape of the main droplet and accounts for accompanying structures. Although these parameters may not provide exact physical information about the droplets, they serve as effective representations or fingerprints of the images, facilitating data analysis.  Before delving into the details, a data distribution analysis was performed to constitute a subset of relevant and well-founded information.

\subsection{Data Distribution and Cleaning}\label{data_cleaning}

\begin{figure}
    \centering
    \includegraphics[width=0.75\textwidth]{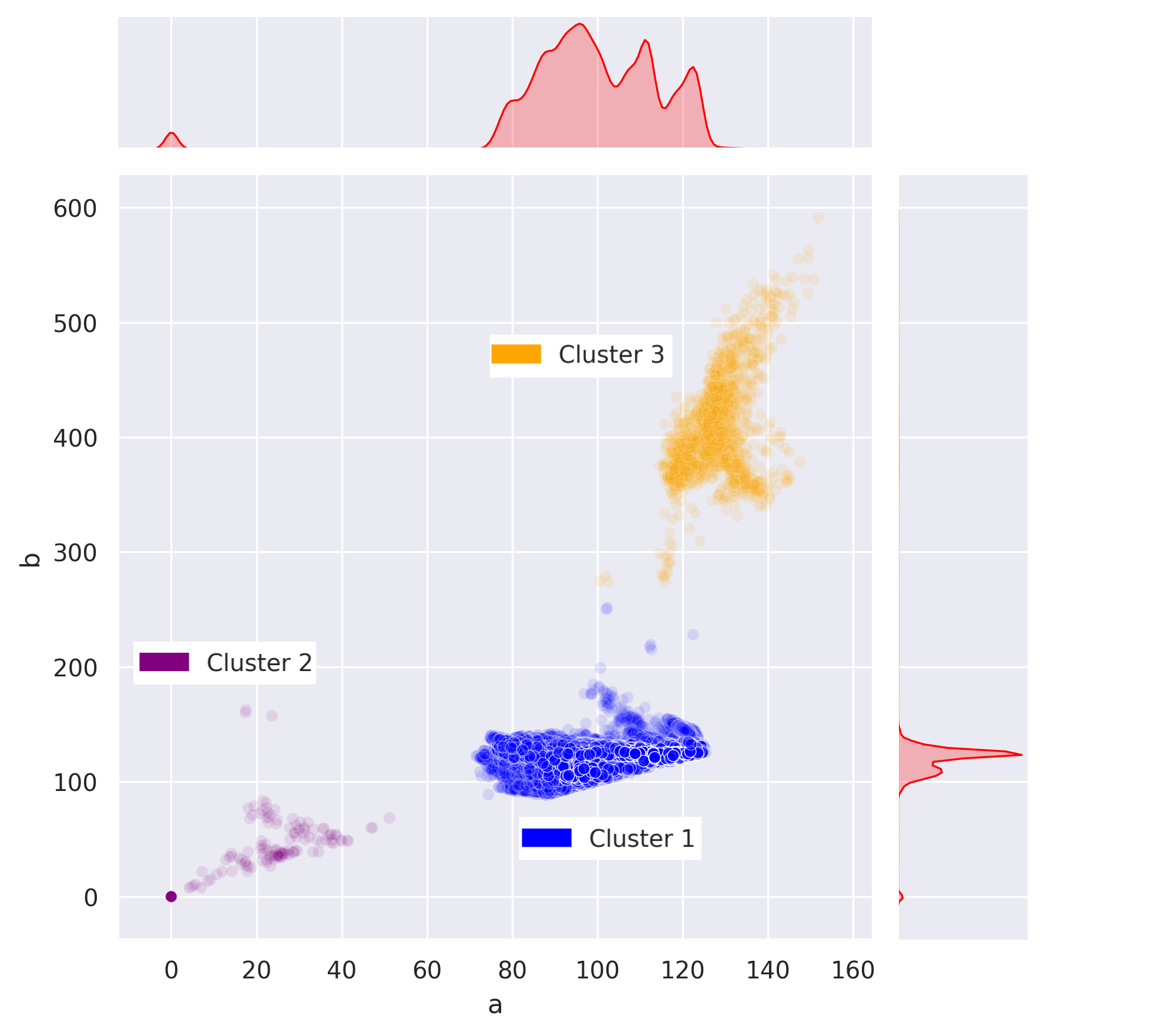}
    \caption{\color{black}{Distribution of semi-axes $a$ and $b$ for detected droplets, with opacity adjusted based on KDE. The top and right graphs show the individual distributions of $a$ and $b$, respectively, using KDE.}}
    \label{fig:a-b_cluster_1}
\end{figure}

\begin{figure}[!htb]
    \centering
    \includegraphics[width=0.8\textwidth]{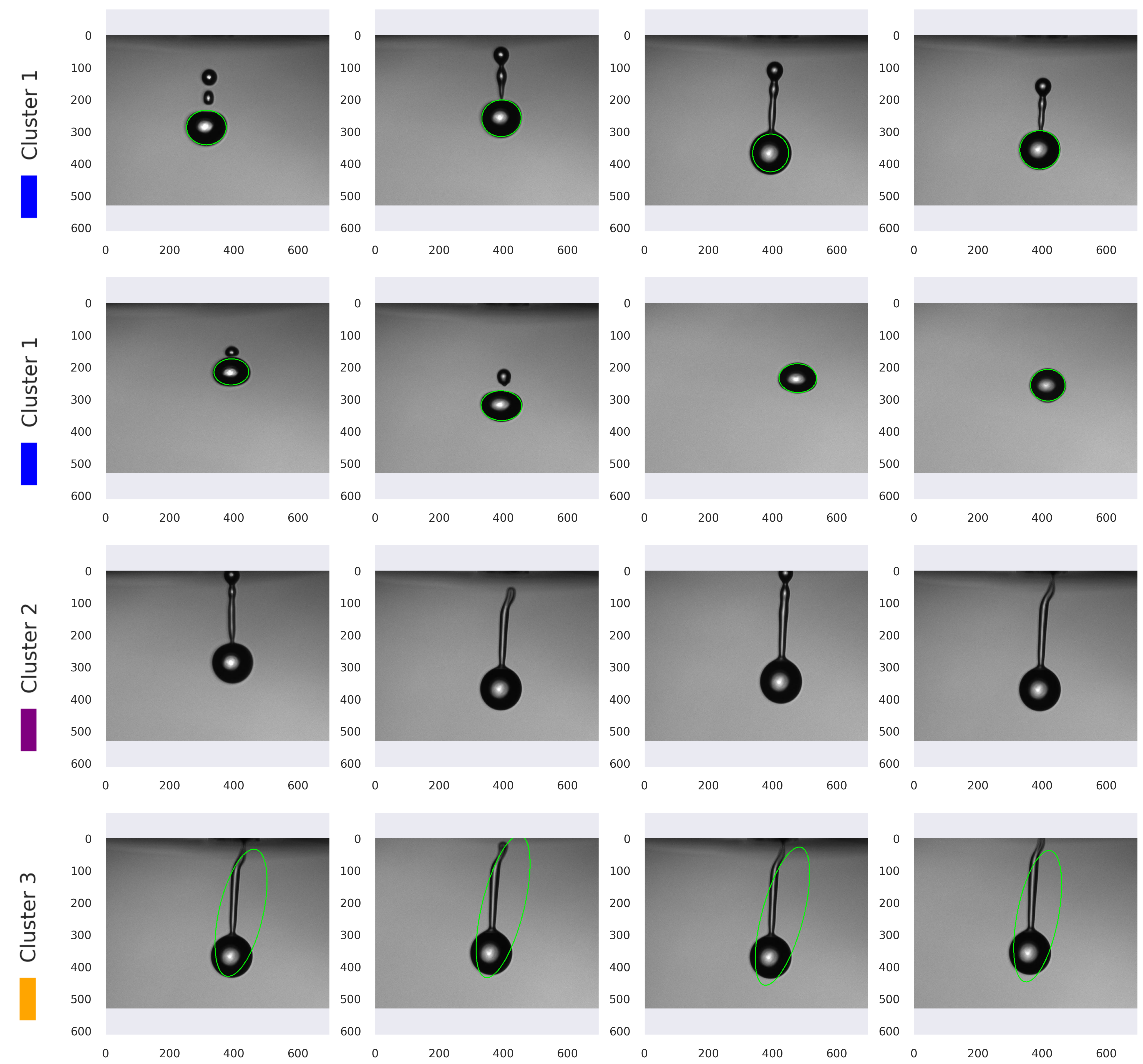}
    \caption{The top two rows show random samples from Cluster 1 of Fig.~\ref{fig:a-b_cluster_1}, where the image processing correctly identifies the geometry of the principal droplet \color{black}(green fitted ellipse) \color{black} in the presence of tails and satellite droplets, respectively. The two lowest rows are randomly selected from the outlier clusters, showing cases where no geometry could be extracted from the images, and instances where the ellipse is incorrectly fitted to the droplet and its tail.}
    \label{fig:a-b_cluster_samples}
\end{figure}

\color{black}Fig.~\ref{fig:a-b_cluster_1} presents a scatter plot of $b$ against $a$, the main descriptors of the shape of the principal droplet, revealing three distinct clusters. These clusters are distinguished by different colors, with opacity levels indicating the probability density estimated using the Kernel Density Estimation (KDE) methodology. Lighter areas represent regions of lower density, while darker areas correspond to denser regions, often associated with outliers or noise.
Cluster 1 accounts for over $80\%$ of the cases and demonstrates successful ellipse fitting to the principal droplet, as illustrated by the random samples in the top two rows of Fig.~\ref{fig:a-b_cluster_samples}. In contrast, the extremes, Clusters 2 and 3, include instances where ellipse fitting faced challenges. These challenges include cases where no ellipse was fitted (third row of Fig.~\ref{fig:a-b_cluster_samples}) or cases where the ellipse enclosed both the droplet and its tail as a single entity (lowest row of Fig.~\ref{fig:a-b_cluster_samples}), likely due to illumination conditions and other factors.\color{black}


To extract a representative subset for further analysis, a specific range of $a$ and $b$ values was selected: $78 \leq a \leq 125$ and $100 \leq b \leq 135$, \color{black}{that aligns with the densest zone of Fig.~\ref{fig:a-b_cluster_1}}. This selection was \color{black}{strengthened} by calculating the first ($Q_1$) and third ($Q_3$) quartiles and using the inter-quartile range (IQR) to remove outliers. This ensures that the subset consists mostly of correctly fitted ellipses, with at least one identified droplet.

\begin{figure}
    \centering
    \includegraphics[width=0.4\textwidth]{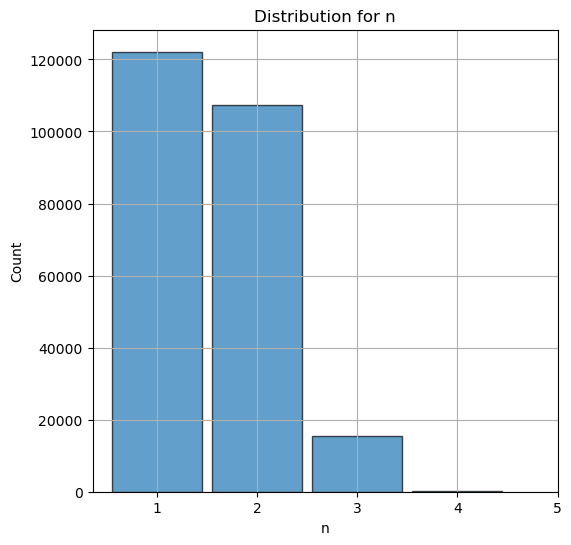}
    \caption{Distribution of the number of detected droplets, $n$.}
    \label{fig:n_dist}
\end{figure}

Based on the established conditions, the cleaned database was obtained, which contains 245,314 data points. ~Fig.~\ref{fig:n_dist} shows the distribution of $n$, indicating that most data is evenly spread across $n=1$ and $n=2$. As previously mentioned, the input parameter ranges were carefully chosen to align with the operational range of the inkjet dispenser, thereby minimizing instances where no droplets are produced ($n=0$). In addition, to maintain data homogeneity, it is logical to consolidate $n$ into two categories: $n=1$ and $n \geq 2$. However, before proceeding with it, it is valuable to analyze the physical insights that can be derived from $n$. 

\subsection{Data Insights on key parameter}
\begin{figure}
    \centering
    \includegraphics[width=0.9\textwidth]{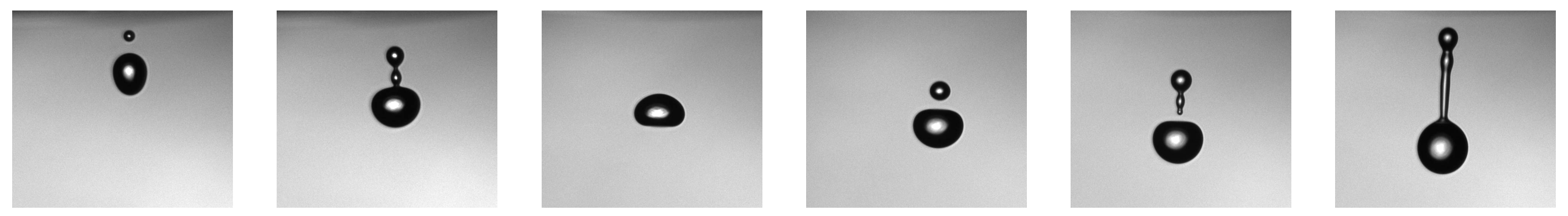}
    \caption{Selected samples representing various cases found in the database.}
    \label{drop_mor}
\end{figure}

After establishing the main conditions for setting up the database, the goal is to gain physical insights into how the operating parameters influence droplet characteristics. To do this, it is essential to describe the properties that can be extracted from the chosen outputs. Fig.~\ref{drop_mor} presents samples representing various cases found in the database. The images clearly show that the main droplet, characterized by $a$ and $b$, varies in both shape and orientation. These droplets range from elliptical shapes, aligned horizontally or vertically, to mushroom-like or circular morphologies, with their characteristics captured by the semi-axis lengths. Additionally, cases involving multiple droplets or droplets with liquid threads can be generally identified by the droplet count, $n$. However, careful consideration is required when using this parameter, as there are instances where $n$ does not accurately reflect the droplet configuration due to the specific nature of the cases. It is also important to differentiate between droplets with liquid threads and those without.

\begin{figure}
    \centering
    \begin{minipage}{0.5\textwidth}
        \centering
        \includegraphics[width=\textwidth]{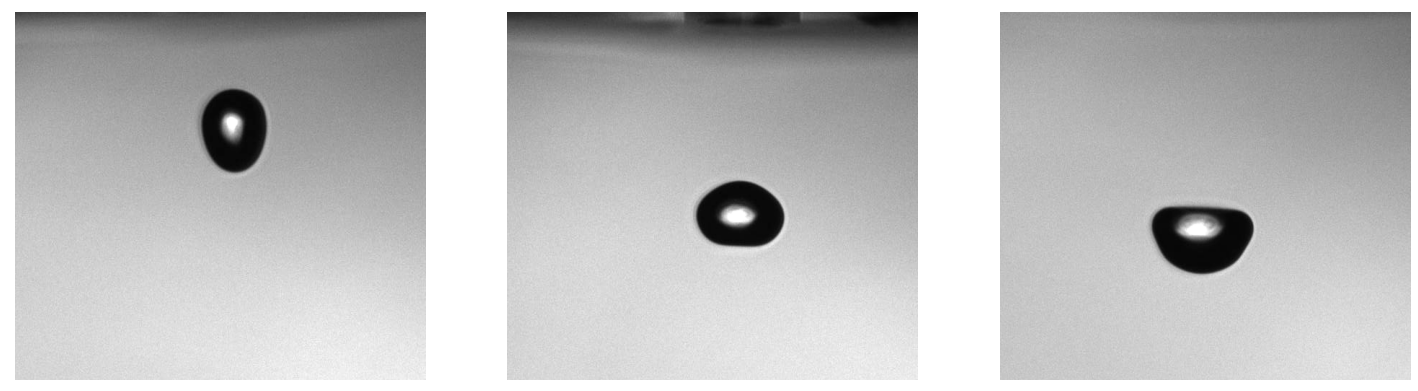}
        \caption*{(a) $n=1$ tagged droplets.}
    \end{minipage}
    \hspace{0.5cm}  
    \begin{minipage}{0.5\textwidth}
        \centering
        \includegraphics[width=\textwidth]{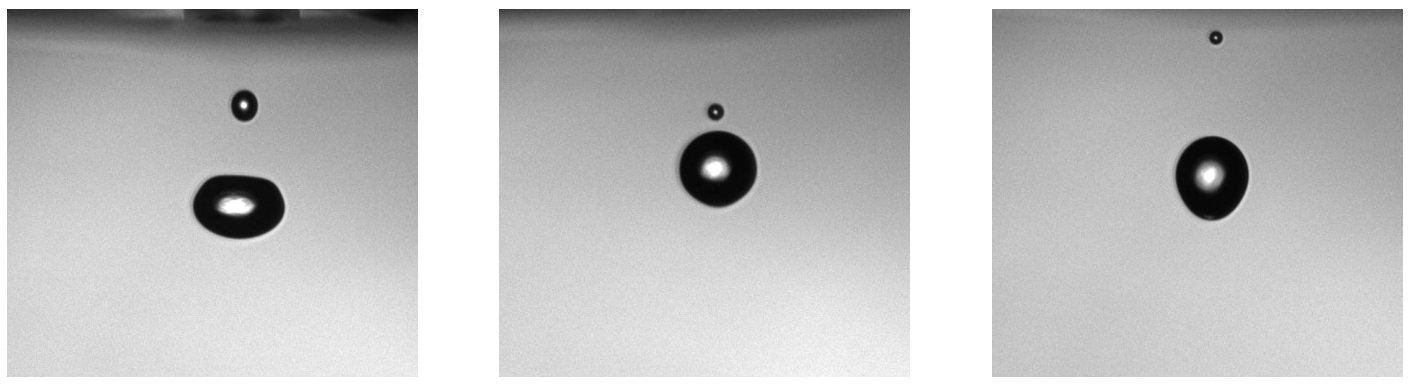}
        \caption*{(b) $n=2$ tagged multi-droplets.}
    \end{minipage}\\
    \begin{minipage}{0.5\textwidth}
        \centering
        \includegraphics[width=\textwidth]{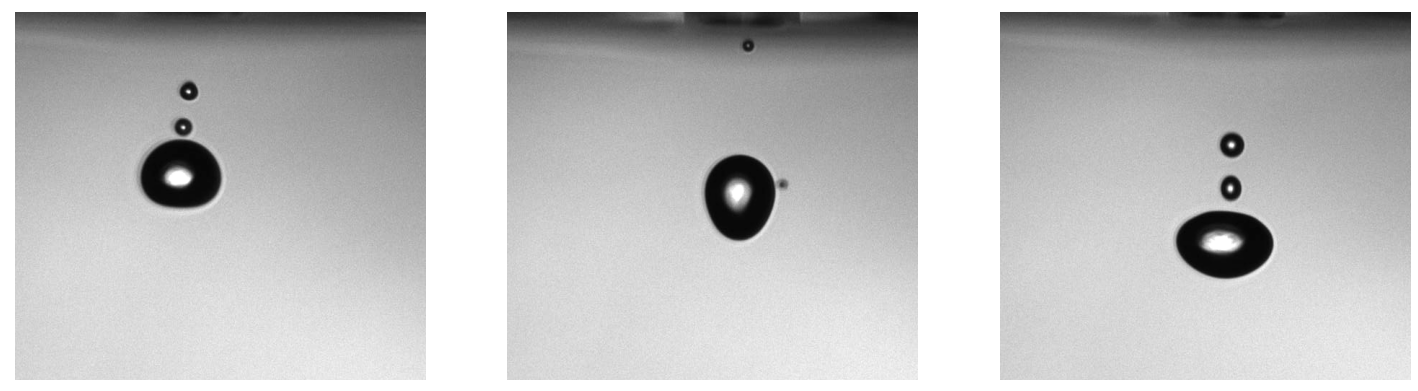}
        \caption*{(c) $n=3$ tagged multi-droplets.}
    \end{minipage}
 \caption{Examples of single and multiple droplet cases.}
  \label{Fig_n}
\end{figure}

\begin{figure}
    \centering
    \includegraphics[width=0.5\textwidth]{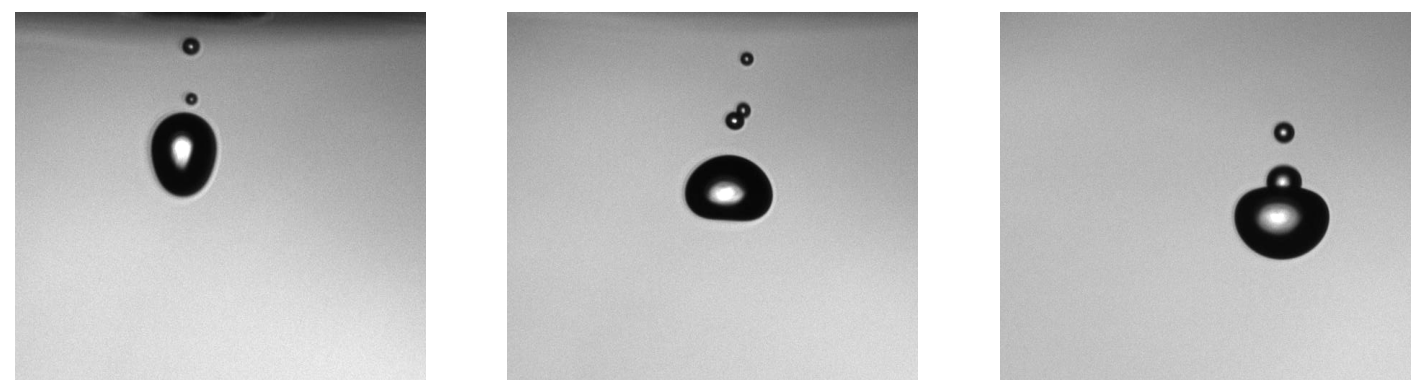}
    \caption{Odd cases tagged with $n=2$.}
    \label{Fig_n_odd}
\end{figure}

Fig.~\ref{Fig_n} shows representative cases where $n$ correctly identifies the number of droplets. However, Fig.~\ref{Fig_n_odd} shows cases labeled as $n=2$ that do not correspond to the droplets in the image. This typically happens either, for instance, when satellite droplets are too small, or far from the main droplet, complicating identification.

\begin{figure}
    \centering
    \begin{minipage}{0.5\textwidth}
        \centering
        \includegraphics[width=\textwidth]{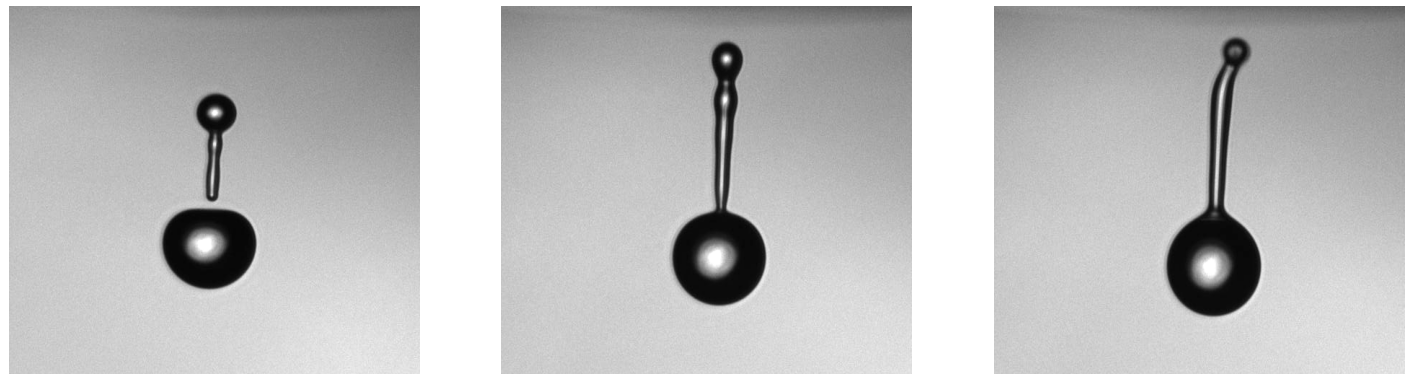}
        \caption*{(a) $n=1$ tagged tailed droplets.}
    \end{minipage}
    \hspace{0.5cm}
    \begin{minipage}{0.5\textwidth}
        \centering
        \includegraphics[width=\textwidth]{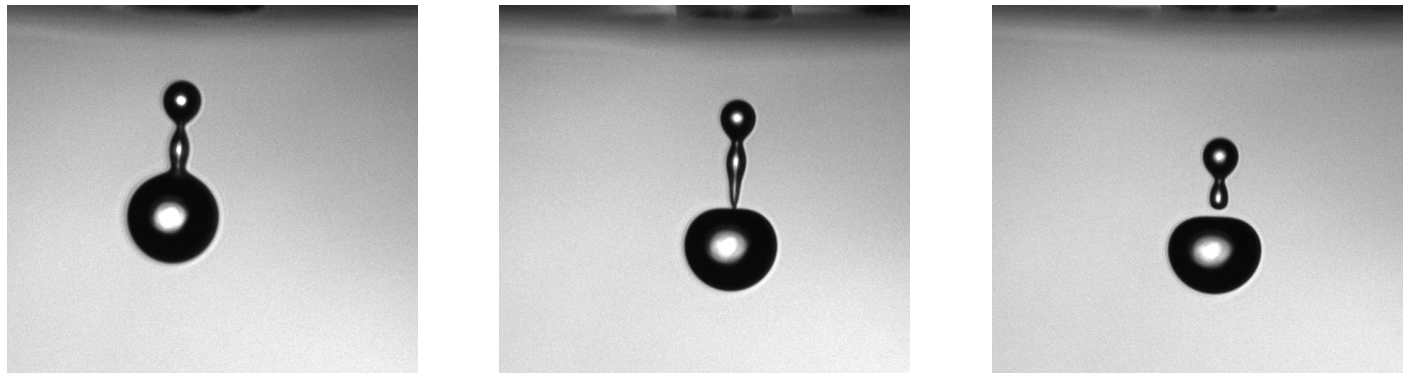}
        \caption*{(b) $n=2$ tagged tailed droplets.}
    \end{minipage}\\
    
    \begin{minipage}{\textwidth}
        \centering
        \includegraphics[width=0.5\textwidth]{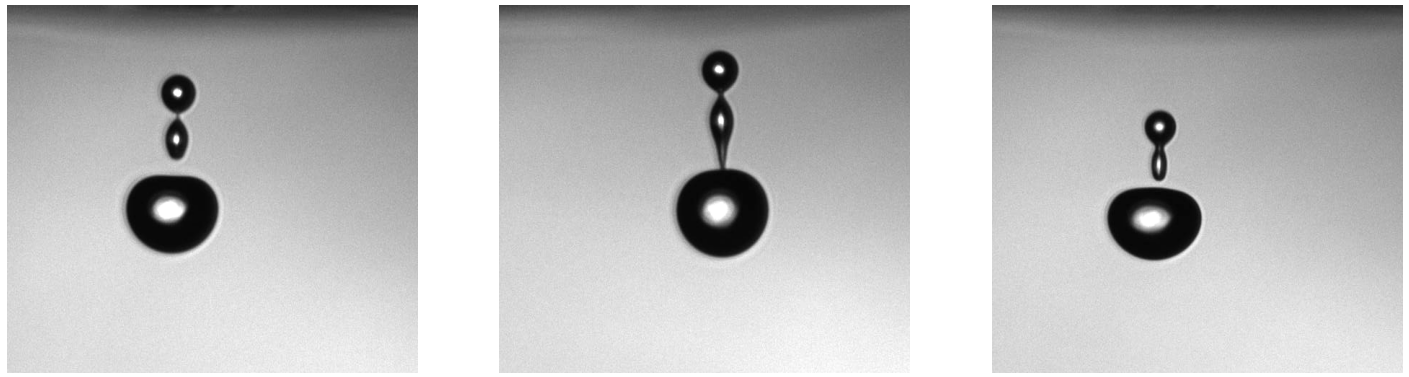}
        \caption*{(c) $n=3$ tagged tailed droplets.}
    \end{minipage}
 \caption{Samples of droplets with liquid threads.}
  \label{Fig_t}
\end{figure}

For droplets with liquid threads or tails, Fig.~\ref{Fig_t} presents representative samples showcasing the diversity of characteristics. 
\color{black}The number of droplets identified by the image processing algorithm is sensitive to the presence of large protuberances and narrow necks in the stream. \color{black} 
For example, in cases with $n=1$, only the main droplet is detected, as the droplet head is significantly larger than the disturbances in the threads. In contrast, for cases with $n=2$, the protuberance at the end of the thread is also recognized as a satellite droplet. This extends to cases with $n=3$, where the thread begins to break into three distinct sections.

Thus, in the case of tailed droplets, the $n$ parameter provides insight into the number of perturbations or breakups along the liquid thread, which will eventually evolve into satellite droplets as the droplet’s time evolution progresses. These stream perturbations can be explained by Plateau-Rayleigh instability or capillary waves \cite{shin_control_2011,mead-hunter_plateau_2012}. Particular attention should be given to specific cases, such as when a spherical droplet pinches off at the end of the liquid thread \cite{dong_visualization_2006}. As the liquid is expelled, the thread elongates, and surface tension pulls the liquid into a spherical shape at the tip. Eventually, this droplet breaks off due to the inertia of the fluid.

From the examples shown, it is evident that the morphology of the main droplet or droplet head is closely related to the presence of satellite droplets or thread disturbances. Therefore, the combined parameters $a$, $b$, and $n$ provide a comprehensive characterization of the droplets, as discussed in the following section. Furthermore, as will be shown, there is a clear correlation between the input voltage parameters and the geometric features of the droplets.

\section{Results and Discussion: A Physical Interpretation}\label{Phys}

Several studies have experimentally investigated the relationship between operational parameters and droplet behavior, typically drawing conclusions based on direct observation \cite{bogy_experimental_1984,malloggi_electrowetting_2008}. However, extracting physical relations across a broad range of operational parameters and deriving meaningful insights from vast amounts of experimental data presents significant challenges. This makes data-driven approaches indispensable for uncovering patterns and correlations in the data~\cite{jang_influence_2009,shin_shape_2014,gong_characterization_2023}.

\begin{figure}[ht]
    \includegraphics[width=\textwidth]{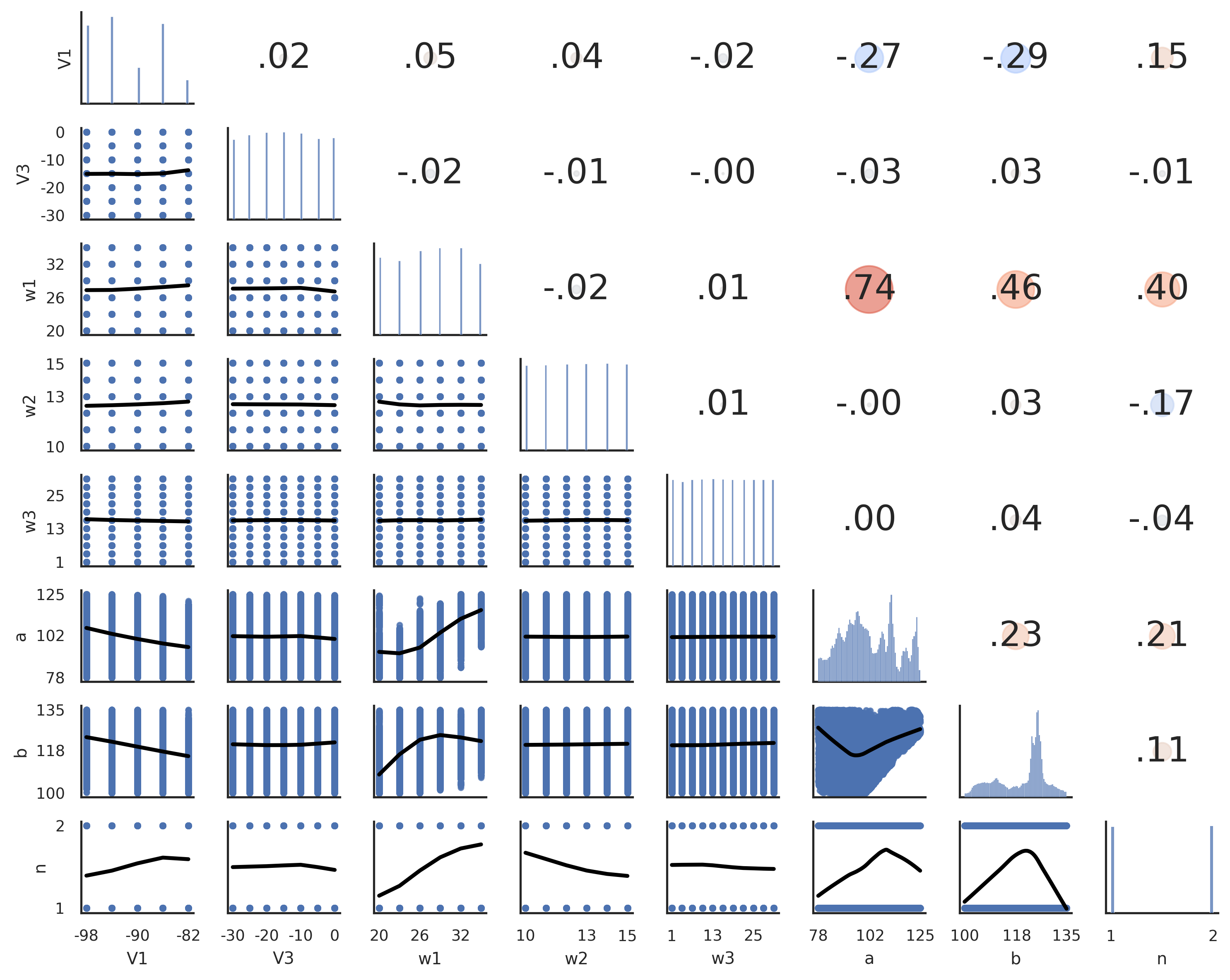}
    \caption{\centering \color{black}{Plot displaying pairwise relationships, distributions, and correlation coefficients between the system variables.}}
    \label{corr}
\end{figure}

\color{black}{To address these challenges, a pair-plot graph is presented in Fig.~\ref{corr} showing pairwise relationships, data distribution and correlation coefficients between selected variables in the data set. The lower triangle of the graph shows scatter plots with fitted regression lines. The diagonal presents histograms for each variable, while the upper triangle includes the Pearson correlation coefficient for each pair of variables. As previously mentioned, the ranges of the second pulse amplitude ($V_2$) and the delays between pulses ($d_1$ and $d_2$) ensuring droplet formation are rather narrow (see Table~\ref{tab:inputs}). Considering that the influence of their variation within the feasible range is negligible, they were not considered in the chart of Fig.~\ref{corr}. Noting that, for the chart, the cases $n\geq{2}$ were grouped together.}

\color{black}{The Pearson correlation coefficients between input and output parameters, provided key insights into how the three-pulse scheme affects droplet characteristics, and reveals relationships that would otherwise be difficult to discern from raw experimental data. This serves as a valuable guide for understanding the relationships between the operational parameters and the geometrical properties of the droplets and allows to find how to systematically adjust the input parameters to obtain the desired outcome.}

\color{black}{From the correlation coefficients and trend curves in Fig.~\ref{corr}, it is clear that the parameters most influencing droplet characteristics are those related to the first pulse and the duration of the second pulse. Additionally, the third pulse plays a crucial role in droplet detachment, acting as a final 'flip' to release the droplet from the nozzle. This pulse takes place after a delay ($d_2$) following the second pulse. Consequently, the timing of its application, governed by $w_2$ and $d_2$, is essential. However, the specific characteristics of the third pulse—its amplitude ($V_3$) and width ($w_3$)—have minimal influence on the droplet, since the detachment occurs almost immediately at the start of this pulse. The last is reveal in the chart by observing the correlation coefficients and noting that the trend curves are practically constant for those cases that involve the parameters ($V_3$) and ($w_3$).}

\begin{figure}[ht]
    \centering
    \begin{minipage}{0.5\textwidth}
        \centering
        \includegraphics[width=\textwidth]{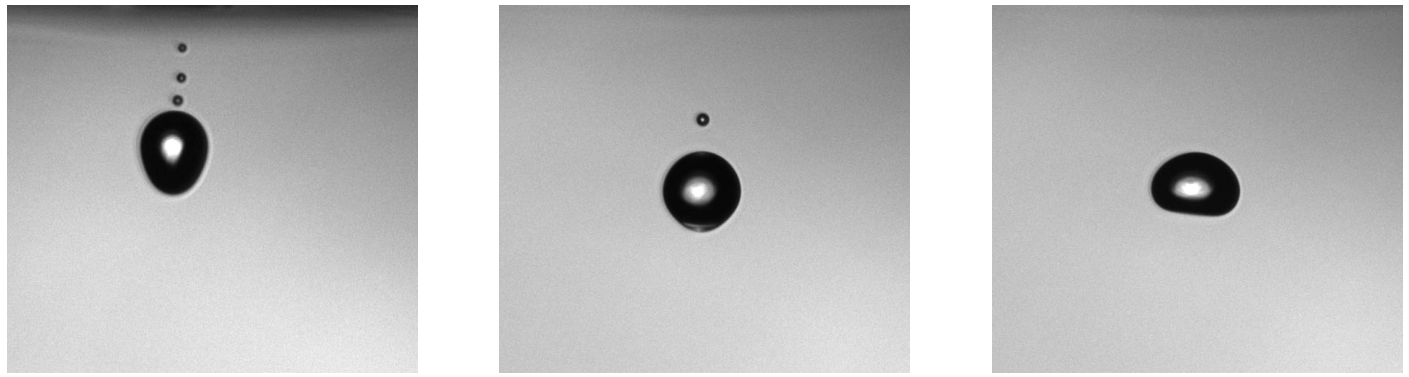}
        \caption*{(a) $w_1=20\mu$s.}
    \end{minipage}
    \hspace{0.5cm}
    \begin{minipage}{0.5\textwidth}
        \centering
        \includegraphics[width=\textwidth]{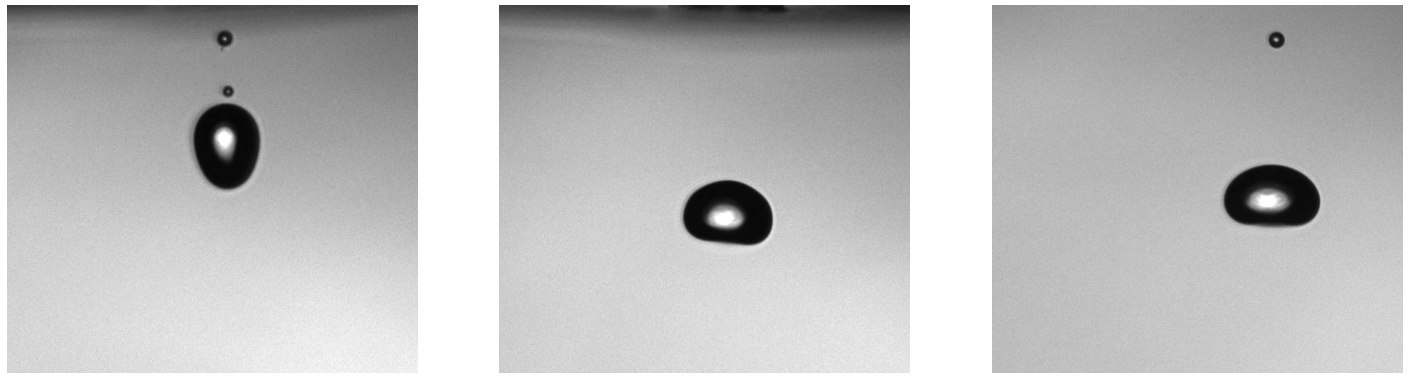}
        \caption*{(b) $w_1=23\mu$s.}
    \end{minipage}\\
    \begin{minipage}{0.5\textwidth}
        \centering
        \includegraphics[width=\textwidth]{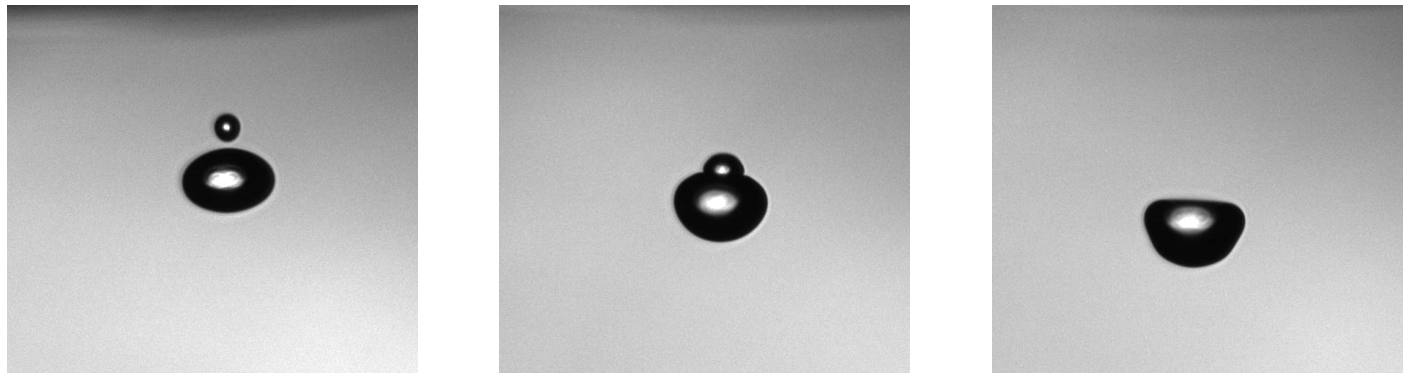}
        \caption*{(c) $w_1=26\mu$s.}
    \end{minipage}
    \hspace{0.5cm}
    \begin{minipage}{0.5\textwidth}
        \centering
        \includegraphics[width=\textwidth]{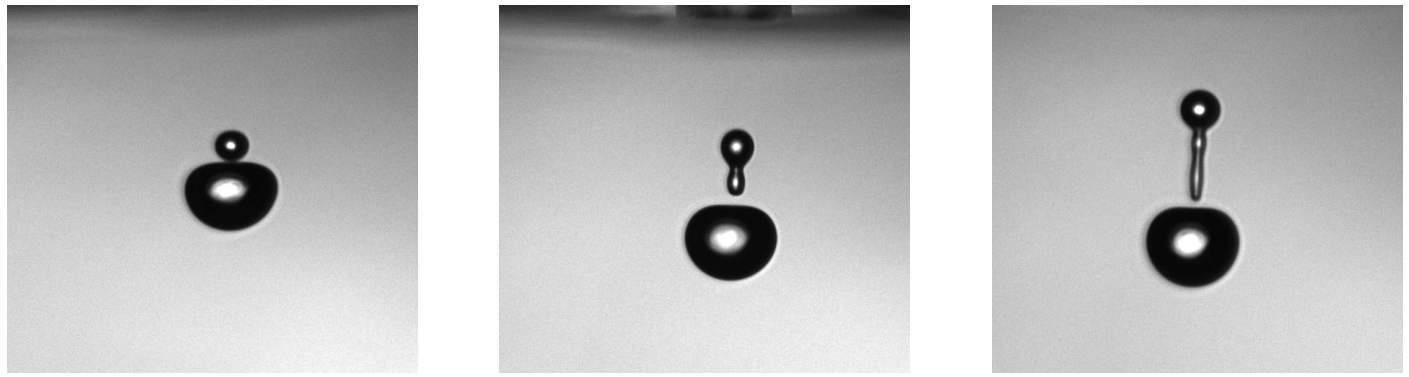}
        \caption*{(d) $w_1=29\mu$s.}
    \end{minipage}\\
    \begin{minipage}{0.5\textwidth}
        \centering
        \includegraphics[width=\textwidth]{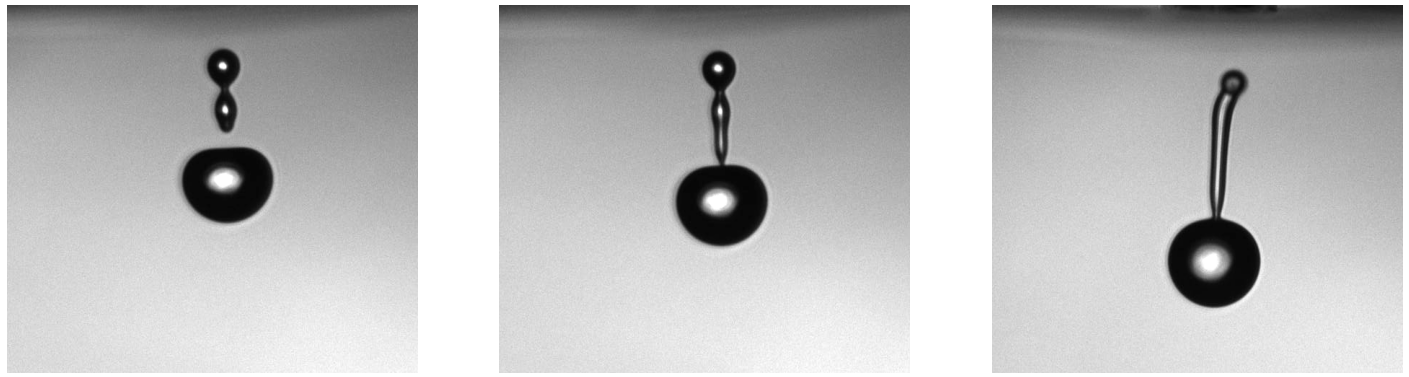}
        \caption*{(e) $w_1=32\mu$s.}
    \end{minipage}
    \hspace{0.5cm}
    \begin{minipage}{0.5\textwidth}
        \centering
        \includegraphics[width=\textwidth]{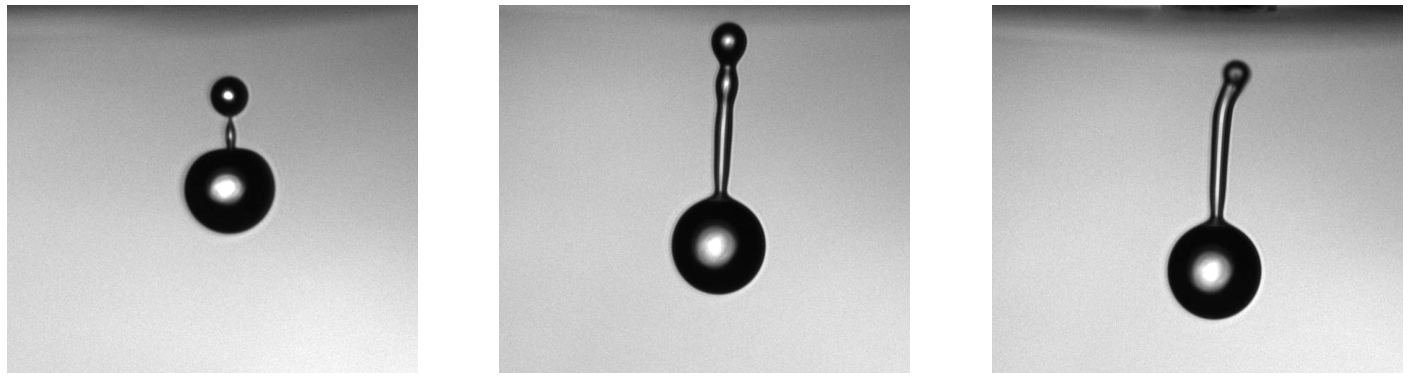}
        \caption*{(f) $w_1=35\mu$s.}
    \end{minipage}
    \caption{Examples of droplet characteristics with varying first pulse duration $w_1$.}
    \label{Fig_w1}
\end{figure}

Among the parameters, the first pulse duration ($w_1$) is the most impactful, particularly on the vertical (jet falling direction) semi-axis length, $a$, indicating that $w_1$ shapes the principal droplet by controlling the amount of expelled ink from the nozzle.
To verify this, Fig.~\ref{Fig_w1} presents samples for different values of $w_1$. As suggested by the \color{black}{chart}, the images show that as $w_1$ increases, the morphology of the main droplet transitions from a single droplet with tiny satellite droplets to multi-droplet formations with larger secondary droplets and liquid threads, culminating in rounder primary droplets with attached liquid threads.

The second important parameters, $V_1$ and $w_2$, influence droplet characteristics in distinct ways. For instance, Fig.~\ref{corr} shows that $w_2$ has a negative correlation with $n$, confirming its role in the formation of satellite droplets and disturbances in the liquid thread, as mentioned previously. In contrast, the influence of the first pulse amplitude ($V_1$) is more subtle and varies depending on the specific case. To exemplify this, we next consider particular cases to illustrate how the most relevant operational parameters ($w_1, V_1$ and $w_2$) influence the droplets' attributes. If the parameter value is not specified, it can assume any value within the bounds of the operating ranges indicated in Table~\ref{tab:inputs}.

\begin{figure}[ht]
    \centering
    \includegraphics[width=0.9\textwidth]{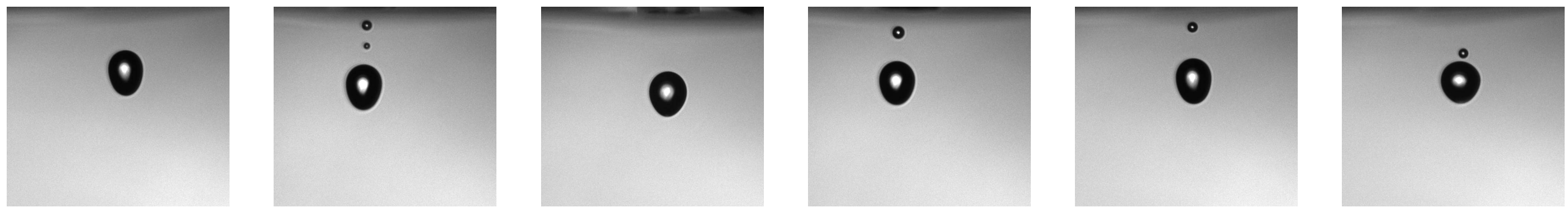}
    \caption{Vertically elongated elliptical-shaped droplets ($w_1=20\mu${s}, $V_{1}={-82}$V or ${-86}$V, and $w_2=15\mu$s).}
    \label{Fig_hor}
\end{figure}

For the sake of illustration, among the collection of cases, six randomly selected samples are provided in Fig.~\ref{Fig_hor}. They were generated with a short first pulse ($w_1 = 20 \mu$s) at low amplitude ($V_1= -82$V or $V_1= -86$V) and a prolonged second pulse ($w_2 = 15 \mu$s), since this configuration pretends to avoid accompanying structures. These elliptical-shaped droplets, elongated along the falling jet axis, are generally accompanied by smaller satellite droplets. The elliptical shape arises because the short and low-amplitude first pulse expels smaller ink volume, preventing long thread formation, while the second pulse stretches the droplet before release, causing oscillations as it falls, alternating between vertical and horizontal orientations \cite{hashemi_enriched_2020}.

\begin{figure}[ht]
    \centering
    \includegraphics[width=0.9\textwidth]{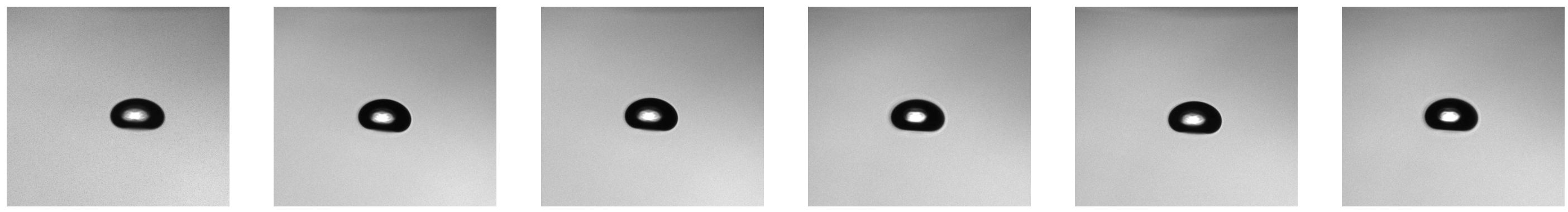}
    \caption{Horizontally elongated elliptical-shaped droplets ($w_1=23\mu${s}, $V_{1}=-98$V, and $w_2=15\mu$s).}
    \label{Fig_ver}
\end{figure}

By slightly increasing the first pulse duration ($w_1=23\mu$s) and raising its amplitude ($V_1=-98$V), while maintaining the second pulse at its maximum duration, horizontally elongated elliptical-shaped droplets without satellite droplets are consistently produced, as shown in Fig.~\ref{Fig_ver}.

As the value of $w_1$ increases, droplet threads become more pronounced, and both the number and size of satellite droplets increase \cite{dong_visualization_2006,shin_control_2011}. The main droplet begins to oscillate once the remnants detach. In cases with larger satellite droplets, the main droplet takes on a more pronounced mushroom-like shape. However, if the tail does not detach, the principal droplet remains nearly circular, indicating that oscillations have not yet started. These cases are associated with longer first pulse durations, specifically $w_1 = 35\mu$s, as confirmed in Fig.\ref{Fig_w1}. Since the first pulse duration is more than twice the maximum value considered for the second pulse, $w_2$ has minimal impact on droplet morphology in these cases. However, the amplitude of $V_1$ strongly affects the characteristics of the liquid threads. To illustrate this, particular values of $V_1$ such as $-98$V and $-86$V, as shown in Table~\ref{tab:inputs}, can be considered.

\begin{figure}[ht]
    \centering
    \includegraphics[width=0.9\textwidth]{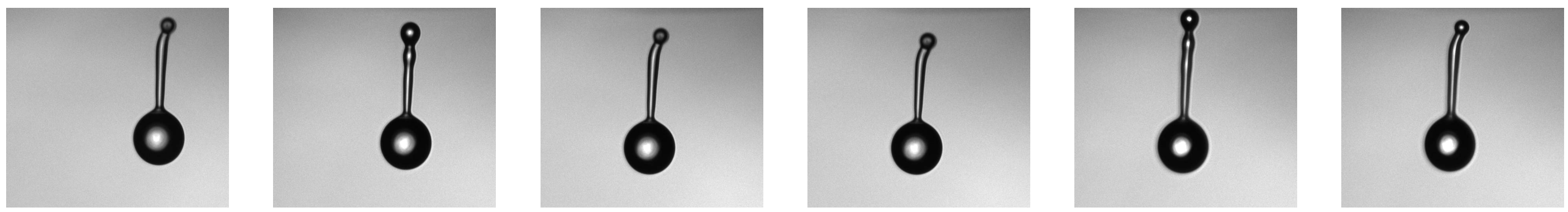}
    \caption{Droplets with long liquid threads ($w_1=35\mu${s} and $V_1=-98$V).}
    \label{Fig_l_t}
\end{figure}

\begin{figure}[ht]
    \centering
    \includegraphics[width=0.9\textwidth]{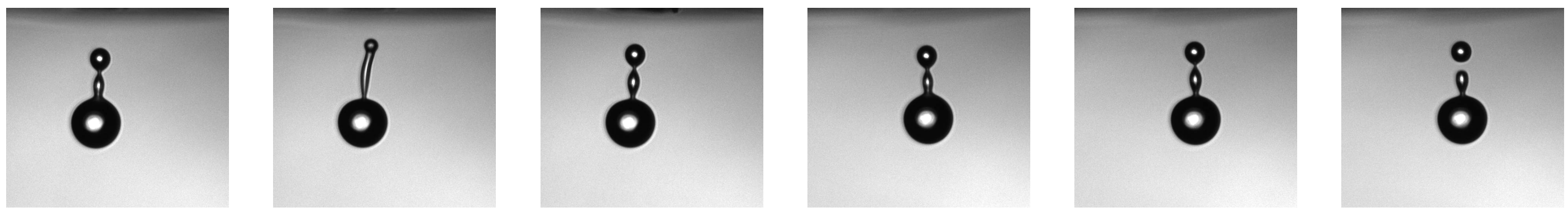}
    \caption{Droplets with shorter liquid threads ($w_1=35\mu${s} and $V_1=-86$V).}
    \label{Fig_s_t}
\end{figure}

When using a long pulse and high amplitude ($V_1 = -98$V), Fig.\ref{Fig_l_t} shows droplets with extended liquid threads and a protuberance at the end, a phenomenon previously described in the literature \cite{dong_visualization_2006}. Although the main droplet initially appears round, it is actually more flattened, as the tail stretches the principal droplet. By reducing the first pulse amplitude to $V_1 = -86$V, the main droplet becomes even rounder (see Fig.\ref{Fig_s_t}), as the shorter tail leads to less stretching and the final disturbance is about to separate.

\begin{figure}
    \centering



    \includegraphics[width=0.4\textwidth]{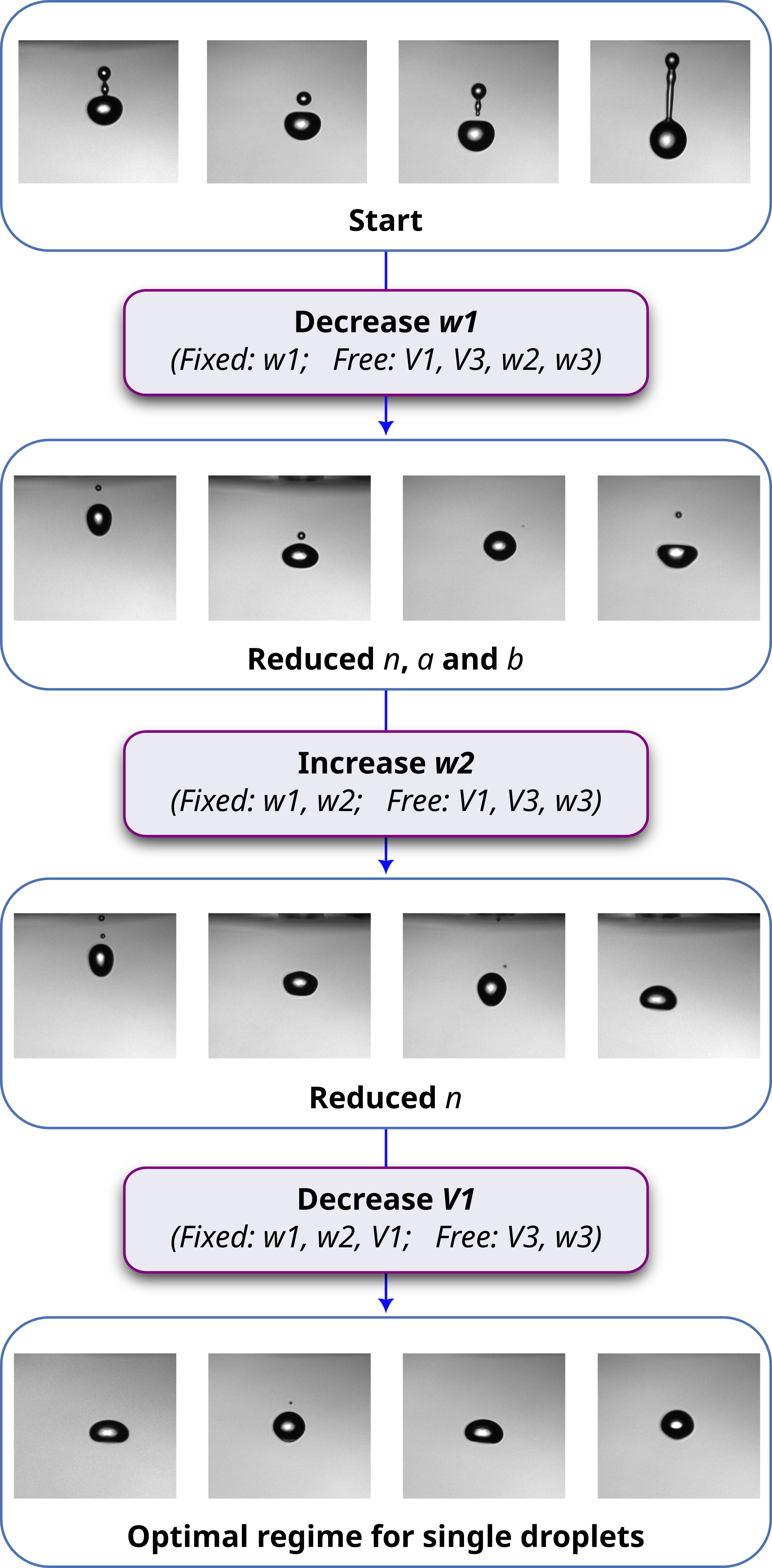}
    \caption{\color{black}Flowchart illustrating the recommended configuration for engineering single droplets by optimizing pulse parameters. At each step, the fixed and free variables are explicitly noted corresponding to those plotted in Fig.~\ref{corr}. Narrow-range variables ($V_2$, $d_1$, and $d_2$) are not explicitly listed but are free to vary within their constrained ranges.\color{black}}
    \label{fig:dig}
\end{figure}

To summarize the key points presented in this section and apply them to a practical scenario, the \color{black}{elements of Fig.~\ref{corr}}  provides valuable guidelines for selecting optimal droplet attributes. In particular, to optimize ink usage—especially with expensive materials—it is essential to eliminate satellite droplets and other remnants \cite{raut_inkjet_2018}. The flow graph in Fig. \ref{fig:dig} illustrates how the dispenser should be configured to consistently produce single droplets. The diagram suggests using a short first pulse with high amplitude, while keeping the second pulse duration at its maximum value.

This recommendation stems from the \color{black}{correlation coefficients and the trend curves}, which shows that a smaller value for $w_1$ results in a smaller vertical semi-axis ($a$), thereby reducing the formation of remnants. This trend is consistent with the examples shown in Fig.~\ref{Fig_w1} and the initial samples in the flow diagram. It is worth noting that for vertically elongated droplets, the images were rotated by the software by 90 degrees, as it always selects the smaller value of the semi-axes as $a$. Along with $a$, the horizontal semi-axis ($b$) and the number of satellite droplets ($n$) are also reduced, with the latter effect being more prominent as the size and number of satellite droplets decrease.

Maximizing the value of $w_2$ further reduces $n$, as expected, compared to other parameter adjustments. The second set of samples in the flow diagram predominantly shows single droplets, with very few cases of multi-droplets, and even in those instances, the satellite droplets are small.

In addition to $w_1$ and $w_2$, the amplitude of the first pulse, $V_1$, also influences droplet characteristics. As indicated by Fig.\ref{corr}, reducing the numerical value of $V_1$ (high amplitude) causes both semi-axis lengths to increase at a similar rate, while $n$ decreases further. By following this approach, most cases result in single droplets, and if satellite droplets are present, they are minimal, as shown in the final image of Fig. \ref{fig:dig}. This observation aligns with the configuration of the operation parameters to get the samples in Fig.~\ref{Fig_ver}, where $w_1$ was specifically set to $23 \mu$s, ending with single droplet cases. While in the flow diagram, $w_1$ was also considered as $20 \mu$s, which results in the droplets with only minor accompanying satellite drops as exhibited in the last samples within the diagram.

\section{Conclusions}\label{Con}
In this study, by characterizing images of droplets, relationships between the morphology of droplets generated by the microdrop dispensing system and the components of the three-voltage pulse scheme, were investigated. Through a data-driven analysis, it was identified that the shape of the principal droplet can be adequately manipulated by adjusting the first and second pulse parameters. These findings complement previous studies, by highlighting that the elliptical envelopes fitted to the main droplets allow to determine whether there are cases of single, multiple or tailed drops. This suggests that with only a few geometric parameters, different attributes of drops can be known, and this has a strong implication for future studies. In this way, other accompanying structures can be guessed from the shape of the primary droplet simply by following the values of the semi-axis lengths. Apart from the examples exhibited in this paper, naturally, something similar could be done to other cases in transition. Overall, the most important insights coming from this work, correspond to the methodologies that can be employed to other equivalent systems when thinking on the applications on droplet engineering to reduce the remnants.  

\backmatter









\section*{Author contribution statement} 
\textbf{Ali R. Hashemi}: Conceptualization, Experiment - design \& execution, Data \& image - processing, Investigation, Software, Writing – original draft. \textbf{Angela M. Ares de Parga-Regalado}: Conceptualization, Data \& image physical insights, Investigation, Methodology, Writing – original draft.  \textbf{Pavel B. Ryzhakov}: Conceptualization, Experiment - design, Funding acquisition, Methodology, Supervision, Writing – review \& editing.  

\section*{Declarations}

\bmhead{Funding} This work was performed in the framework of DIDRO project (Towards establishing a Digital twin for manufacturing via drop-on-demand inkjet printing, \textit{Proyectos Estratégicos Orientados a la Transición Ecológica y a la Transición Digital 2022-2024}, reference TED2021-130471B-I00) financed by Ministry of Spain MICIU/AEI10.13039/501100011033 and by European Union Next GenerationEU/ PRTR, and DECIMA project, PID2022-137472OB-I00 financed by MICIU/AEI/10.13039/501100011033/FEDER, UE.

\bmhead{Availability of data and materials} The dataset of captured images and the dataset of extracted parameters are openly available under a CC BY-NC-SA license at \url{https://doi.org/10.5281/zenodo.13862494} (Ref.~\cite{hashemi_inkjet_2024}). The code for driving the inkjet dispenser, capturing, processing images, and extracting features is published at \url{https://github.com/DropletDynamics/InkJetDroplet} (Ref.~\cite{hashemi_inkjetdroplet_2024}).

\bmhead{Ethics approval and consent to participate} Not applicable 

\bmhead{Consent for publication} Not applicable 

\bmhead{Conflict of interest} There is no conflict of interest regarding this work

\end{document}